 \documentclass[preprint,prc,showpacs,preprintnumbers,
                superscriptaddress,floatfix]{revtex4}
\usepackage{graphicx,epsfig,latexsym,amssymb}
\usepackage[compact,tiny]{titlesec}
\usepackage{multirow,amsmath}
\usepackage{dcolumn,fancyhdr,makeidx}
\usepackage{subfigure,color}
\usepackage[bookmarksnumbered,plainpages]{hyperref}
\newcommand{\beq}{\begin{equation}}
\newcommand{\eeq}{\end{equation}}
\newcommand{\beqn}{\begin{eqnarray}}
\newcommand{\eeqn}{\end{eqnarray}}
\newcommand{\bsub}{\begin{subequations}}
\newcommand{\esub}{\end{subequations}}
\newcommand{\bpm}{\begin{pmatrix}}
\newcommand{\epm}{\end{pmatrix}}

\newcommand{\svec}[1]{{\mbox{\boldmath$#1$}}}

\newcommand{\balp}{\mbox{\boldmath$\alpha$}}

\newcommand{\bgam}{\mbox{\boldmath$\gamma$}}

\newcommand{\bj}{{\mathbf{j}}}
\newcommand{\bk}{{\mathbf{k}}}

\newcommand{\bp}{{\mathbf{p}}}

\newcommand{\br}{{\mathbf{r}}}

\newcommand{\bt}{{\mathbf{t}}}

\newcommand{\bV}{{\mathbf{V}}}

\newcommand{\dfra}{\displaystyle\frac}
\begin{document}
 \title{Core Polarization and Tensor Coupling Effects on Magnetic Moments of Hypernuclei }
 \author{Yao Jiang-Ming }
 \affiliation{School of Phyics, and SK Lab. Nucl. Phys. $\&$ Tech.,
 Peking University, Beijing 100871, China}
\author{L\"{u} Hong-Feng }
\affiliation{College of Science, Chinese Agriculture University,
Beijing 100083, China}
\author{Hillhouse Greg}
\affiliation{School of Phyics, and SK Lab. Nucl. Phys. $\&$ Tech.,
Peking University, Beijing 100871, China}
 \affiliation{Department of
Physics, University of Stellenbosch, Stellenbosch, South Africa}

\author{Meng Jie }\thanks{E-mail: mengj@pku.edu.cn}
\affiliation{School of Phyics, and SK Lab. Nucl. Phys. $\&$ Tech.,
Peking University, Beijing 100871, China} \affiliation{Department of
Physics, University of Stellenbosch, Stellenbosch, South Africa}
\affiliation{Institute of Theoretical Physics, Chinese Academy of
Science, Beijing 100080}
\affiliation{Center of Theoretical Nuclear Physics, National Laboratory of \\
       Heavy Ion Accelerator, Lanzhou 730000}

\begin{abstract}
 The effects of core polarization and tensor coupling  on the
 magnetic moments in $^{13}_\Lambda$C, $^{17}_\Lambda$O, and
 $^{41}_\Lambda$Ca $\Lambda$-hypernuclei are studied in the Dirac
 equation with scalar, vector and tensor potentials. It is found that
 the effect of core polarization on the magnetic moments is suppressed
 by $\Lambda$ tensor coupling. The $\Lambda$ tensor potential reduces
 the spin-orbit splitting of $p_\Lambda$ states considerably. However,
 almost the same magnetic moments are obtained using the hyperon wave
 function obtained via the Dirac equation either with or without the
 $\Lambda$ tensor potential in the electromagnetic current vertex.
 The deviations of magnetic moments for $p_\Lambda$ states from the
 Schmidt values are found to increase with nuclear mass number.
\end{abstract}
 \maketitle

 Since the first discovery of $\Lambda$ hypernuclei by observing
 cosmic-rays in emulsion chambers~\cite{Danysz53},
 hypernuclei -- which are nuclei with one or more of the nucleons replaced
 with hyperons -- have been used as a natural laboratory to study
 hyperon-nucleon and hyperon-hyperon
 interactions~\cite{Chrien89,Dover89,Bando90,Lu04,Hashimoto06}.

 The magnetic moments of hypernuclei are important physics
 observables, since they are sensitive to spin and angular momentum
 structure as well as spin-dependent hyperon-nucleon interactions.
 In particular, these quantities provide direct information
 about the properties of hadrons in the nuclear medium.
 The first study of hypernuclear magnetic moments was performed
 for light p-shell hypernuclei within a three-body cluster model~\cite{Motoba85}.
 Thereafter, magnetic moments of light hypernuclei were theoretically
 studied within the shell model, and predicted to be
 around the Schmidt lines~\cite{Tanaka89}. However, the Schmidt values are
 obtained from extreme single-particle model neglecting the core
 polarization due to a particle or a hole added.
 Therefore, magnetic moments of hypernuclei may differ from the Schmidt
 values, especially if exchange currents from heavier mesons or other
 exotic phenomena are considered~\cite{Saito97}.

 Relativistic mean-field (RMF) theory has been successfully
 applied in the analysis of nuclear structure for nuclei ranging
 from light to superheavy elements~\cite{Serot86,Reinhard89,Ring96,Vretenar05,Meng06},
 as well as for studies of the magnetic moments of doubly closed shell nuclei plus or minus
 one nucleon~\cite{Furnstahl87,Hofmann88,Yao06}.
 Hence, it is natural to apply RMF theory to study the properties of
 hypernuclei~\cite{Boguta81,Rufa87,Mares94,Vretenar98,Lu03},
 and in particular hypernuclear magnetic moments.

 In the relativistic approach, the magnetic moments arise from a compensation
 of two effects, namely, the enhancement of the valence charged particle current due to
 the reduction of the nucleon mass and the contribution of
 an additional current from polarized core nucleons~\cite{McNeil86,Furnstahl87,Hofmann88,Yao06}.
 However, this cancelation is not expected in a $\Lambda$-hypernucleus
 due to the charge neutrality of a $\Lambda$ hyperon. Therefore,
 the polarized proton current induced by
 a valence hyperon causes the total magnetic moment to deviate from
 the Schmidt value. Such deviation has even been suggested as an
 indicator of relativistic effects in nuclei~\cite{Cohen87,Mares90}.
 However, these relativistic calculations omitted the tensor coupling of the
 vector field to the $\Lambda$~\cite{Jennings90}.
 The inclusion of a strong $\Lambda$ tensor coupling will renormalize
 the electromagnetic current vertex in the nuclear medium, and bring
 the magnetic moment of $\Lambda$-hypernucleus with
 $\ell_\Lambda=0$ close to the Schmidt value, though not for the case
 of $\ell_\Lambda\ne0$~\cite{Gattone91}. In addition, the $\Lambda$ tensor
 coupling will give rise to a tensor potential which is very important for
 the explanation of the small $\Lambda$ hypernucleus spin-orbit
 interaction~\cite{Noble80,Jennings90}.

 In view of these facts, in this letter, we study the effects of
 core polarization and tensor coupling on the magnetic moments of $\Lambda$ hypernucleus
 within a simple relativistic model with scalar, vector potentials and for the first time
 we also consider tensor potential.

 The Dirac equation for baryons~($B=n,p,\Lambda$) with a tensor potential is given by,
 \beqn
 \label{Dirac}
  \left[i\gamma_\mu\partial^\mu-(M_B+S_B(\br))
       - \gamma_\mu V^\mu_B(\br)
       +\frac{f_{Bv}}{g_{Bv}}\frac{1}{2M_B}\sigma_{\nu\mu}\partial^\nu V^\mu_B(\br)\right]
       \psi_a(\br) = 0
 \eeqn
 where $\sigma_{\nu\mu}=\frac{i}{2}[\gamma_\nu,\gamma_\mu]$,
 $M_B$ is the baryon mass, $S_B(\br)$ and $V^\mu_B(\br)$ are the
 corresponding scalar and vector potentials.
 For spherical nuclei, the space-like components of the vector potential
 $\bV_B(\br)$ vanish due to time-reversal invariance.
 The Dirac spinor $\psi_\alpha(\br)$ ($\alpha = \{n, \kappa, m, \tau\}$) is
 characterized by quantum numbers $\kappa(=(\ell-j)(2j+1))$, $m, \tau$ and
 $n$ -- where $\ell$ and $j$ represent the orbital and total angular momenta respectively,
 $m$ is the projection of $j$ on the $z$-axis, $\tau$ denotes isospin and $n$ refers
 to the radial quantum number -- and is given by,
 \beq
  \label{eq:SRHspinor}
   \psi_{\alpha}(\svec r)
    = \bpm i \dfrac{G_{n\kappa}(r)}{r} \\[1em]
             \dfrac{F_{n\kappa}(r)}{r}\svec\sigma\cdot\hat{\svec r}
      \epm
      Y_{jm}^{\ell}(\theta,\phi)\chi_{\tau},
 \eeq
 where $G_{n\kappa}(r) / r$ and $F_{n\kappa}(r) / r$ are radial wave
 functions for the upper and lower components and $ Y^{\ell}_{jm}(\theta,\phi)$
 denotes the conventional spinor spherical harmonics.
 Substituting Eq.~(\ref{eq:SRHspinor}) into Eq.~(\ref{Dirac})
 yields the following radial Dirac equations
 \bsub
 \label{eq:SRHDirac}
 \beqn
 \epsilon_\alpha G_{n\kappa}
    & = & \left( -\dfrac{d}{d r} + \dfrac{\kappa}{r} + U_B^T(r)\right) F_{n\kappa}
        + \left( M + S_B(r) + V^0_B(r) \right) G_{n\kappa} ,    \\
 \epsilon_\alpha F_{n\kappa}
    & = & \left( +\dfrac{d}{d r} + \dfrac{\kappa}{r} + U_B^T(r)\right) G_{n\kappa}
        - \left( M + S_B(r) - V^0_B(r) \right) F_{n\kappa},
 \eeqn
 \esub
 with the tensor potential defined by
 $U^T_B(r)=\dfra{f_{Bv}}{g_{Bv}}\frac{1}{2M_B}\partial_r V^0_B(r)$.

 In a $\Lambda$ hypernucleus, due to the charge neutrality of a $\Lambda$
 hyperon, the Dirac $\Lambda$ current has no contribution to the
 Dirac magnetic moment, therefore the Dirac magnetic moment in
 $\Lambda$ hypernucleus is only given by polarization currents from
 the nuclear core. The polarized electromagnetic current
 $\bj^{\rm em}_c$ can be approximated by summing the contributions from the
 perturbation due to the inclusion of $\Lambda$ to all orders in
 symmetric nuclear matter as~\cite{Furnstahl87,Cohen87},
 \beqn
  \label{PolarizedC}
  \bj^{\rm em}_c
    =\frac{1}{2}\sum^{k_F}_{k,\sigma}\psi^\dagger_N (\bk,\sigma)\balp \psi_N(\bk,\sigma)
   &=&-\frac{1}{2}\frac{g_{\Lambda v}}{g_{Nv}}\psi^\dagger_\Lambda (\bt,\sigma)\balp \psi_\Lambda(\bt,\sigma)
      [1+\frac{E^\ast_{k_F}}{\lambda_{Nv}\rho_N}]^{-1},
 \eeqn
 where $\lambda_{Nv}=g^2_{Nv}/m^2_v$, $\bk, \sigma$ and $\bt, \sigma$ are respectively
 the momenta and spin of the nucleon and the hyperon. The Fermi momentum $k_F$ is given by the
 density distribution of the nuclear core $\rho_N(r)$,
 $k_F=[\dfra{3\pi^2}{2}\rho_N(r)]^{1/3}$, and the Fermi energy
 is~$E^\ast_{k_F} =  \sqrt{k^2_F+ M^{\ast2}_N}$,
 with the scalar mass given by~$M^{\ast}_N=M_N+S(r)$~\cite{Long06}.

 The Dirac magnetic moment corresponding to the current in Eq.~(\ref{PolarizedC}) is,
 \beqn
  \label{DiracMM}
   \mu_D
   =-\frac{1}{2}\frac{\kappa j}{(\kappa-\frac{1}{2})(\kappa+\frac{1}{2})}
    (\frac{2M_\Lambda c^2}{\hbar c})\int dr  rF_{n\kappa}(r)G_{n\kappa}(r)B_f(r),
 \eeqn
 where the reduction factor $B_f(r)$ is due to the core
 polarization,
 \beq
  \label{Bfunction}
   B_f(r) =
           \frac{g_{\Lambda v}}{g_{Nv}}[1+\frac{E^\ast_{k_F}}{\lambda_{Nv}\rho_N}]^{-1}
           \frac{M_N}{M_\Lambda}.
 \eeq
 According to the Gordon identity,
 $ i\bar\psi_\Lambda\sigma^{\mu\nu}\psi_\Lambda \dfra{(p+p^\prime)_\nu}{2M_\Lambda}
 =\bar\psi_\Lambda\gamma^\mu\psi_\Lambda-\bar\psi_\Lambda \psi_\Lambda
 \dfra{(p+p^\prime)^\mu}{2M_\Lambda}$, with $p$ and $p^\prime$ being
 the momenta of the initial and final $\Lambda$~\cite{Bjorken64},
 the inclusion of a tensor coupling $f_{\Lambda v}/g_{\Lambda v}=1$
 will modify the vertex $\bgam$ to be $(\bp+\bp^\prime)/2M_\Lambda$.
 After partly being canceled, the core polarized electromagnetic current
 in Eq.~(\ref{PolarizedC}) is left with the core polarized convection
 current~\cite{Gattone91},
 \beqn
  \label{PolarizedCT}
   \bj^{\rm em}_c
   &=&-\frac{g_{\Lambda v}}{g_{Nv}}\frac{1}{2M_\Lambda} \bar\psi_\Lambda (\bt,\sigma)
       \frac{\nabla}{i}\psi_\Lambda
       (\bt,\sigma)[1+\frac{E^\ast_{k_F}}{\lambda_{Nv}\rho_N}]^{-1}.
 \eeqn

 Therefore the inclusion of a tensor coupling modifies the Dirac magnetic
 moment in Eq.(\ref{DiracMM}) as,
 \beqn
 \label{Tensor:DiracMM}
  \mu_D &=& -\frac{j}{2} \int r^2dr
             \left[G^2_{n\kappa}(r)-F^2_{n\kappa}(r)
            -\frac{\Omega_\kappa}{2\ell_\kappa+1}G^2_{n\kappa}(r)
            -\frac{\Omega_\kappa}{2\ell_{-\kappa}+1}F^2_{n\kappa}(r)\right]B_f(r),
 \eeqn
 where $\Omega_\kappa=1~(-1),~\ell_\kappa=-\kappa-1~(\kappa)$ for $\kappa<0~(>0)$.

 In contrast, the tensor coupling and core polarization
 effects do not modify the anomalous magnetic moment explicitly~\cite{Gattone91,Yao06-NP},
 \beqn
 \mu_{\rm a} &=& 2\mu_B j\Omega_\kappa \int r^2dr
 \left[\frac{G^2_{n\kappa}(r)}{2\ell_\kappa+1}+\frac{F^2_{n\kappa}(r)}{2\ell_{-\kappa+1}}\right],
 \eeqn
 where the free anomalous gyromagnetic ratio $\mu_B$ are respectively: $\mu_p=1.793$,
 $\mu_n=-1.913$, and $\mu_\Lambda=-0.613$ in nuclear magnetons (n.m).

 As Woods-Saxon potentials are reliable and often adopted for the description of
 nucleus, the scalar and vector potentials in Eq.~(\ref{eq:SRHDirac}) are chosen
 to be Woods-Saxon forms in order to examine the tensor coupling and core polarization
 effects,
 \bsub\beqn
  V^0_B(r)+S_B(r) &=& \frac{U_0}{1+e^{(r-R)/a}},\\
  V^0_B(r)-S_B(r) &=& \frac{-\lambda_{so} U_0}{1+e^{(r-R)/a}},
 \eeqn\esub
 with the potential depth $U_0=-46.4$~MeV, diffusion parameter $a=0.6$~fm,
 $R=[1.19-0.45A^{-2/3}] A^{1/3}$, and $\lambda_{so}=15.99$ according to Ref.~\cite{Gattone91}.
 The masses of the baryons and vector field are respectively,
 $M_n=M_p=939.0$~MeV,  $M_\Lambda=1115.6$~MeV, and $m_v=784$~MeV.
 The vector coupling constant is chosen as $g_{Nv}= 13.0$~\cite{Long04} and
 $g_{\Lambda v}/g_{Nv}= 2/3$ according to a naive quark model. In addition,
 tensor coupling between nucleons has been shown to be small~\cite{Cohen91},
 and hence the effect thereof is neglected in this paper, i.e., $U_N^T(r)=0$.

 The coupled differential equations~(\ref{eq:SRHDirac}) are solved using the shooting
 method~\cite{Horowitz81} combined with Runge-Kutta algorithms and
 employing appropriate boundary conditions in a spherical box of radius $R = 20$
 fm with a step size of $0.1$ fm. The single-particle energies for
 $s_\Lambda$ states in $^{13}_\Lambda$C, $^{17}_\Lambda$O, and $^{41}_\Lambda$Ca are
 $-11.1, -13.1$, and $-19.2$~MeV respectively.
 The effect of the tensor potential $U^T_\Lambda(r)$ is to reduce the spin-orbit
 splitting of $p_\Lambda$ states from $1.15, 1.40$, and $1.10$~MeV to $0.19, 0.25$,
 and $0.23$~MeV respectively.
 The results are in agreement with the observed $\Lambda$ binding energy -11.7 MeV,
 and the splitting size in $p_\Lambda$ states $152\pm54 \pm36$~keV~\cite{Ajimura01}
 in $^{13}_\Lambda$C.

 Table~\ref{tab1} displays the total magnetic moment $\mu$ for the $\Lambda$-hypernuclei
 $^{13}_\Lambda$C, $^{17}_\Lambda$O, $^{41}_\Lambda$Ca, the
 summation of the corresponding polarized Dirac magnetic moment $\mu_D$ and the
 anomalous magnetic moment $\mu_a$.
 The Schmidt values are given in the first row; those with ``Relativistic''
 are obtained according to Eq.~(\ref{PolarizedC}) without tensor coupling,
 i.e., tensor potential $U^T_\Lambda(r)=0$;
 those with ``Tensor*'' and ``Tensor'' are obtained according to
 Eq.~(\ref{PolarizedCT}), but using the hyperon wave function
 in the Dirac equation either without or with the $\Lambda$ tensor
 potential.

  \begin{table}[h!]
   \centering
   \tabcolsep=6pt
   \caption{Magnetic moments of $\Lambda$-hypernuclei $^{13}_\Lambda$C, $^{17}_\Lambda$O,
   $^{41}_\Lambda$Ca in units of nuclear magnetons (n.m), where the Schmidt values are given
   in the first row; those with ``Relativistic'' are obtained
   without tensor coupling; those with ``Tensor*'' and ``Tensor'' are obtained
   according to Eq.~(\ref{PolarizedCT}), but using the hyperon wave function
   in the Dirac equation either without or with the $\Lambda$
   tensor potential. }
   \begin{tabular}{ c |c c c|ccc|ccc}
   \hline
   \hline
   \multirow{2}*{$\mu$ (n.m.)}   & \multicolumn{3}{c|}{$^{13}_\Lambda$C} &   \multicolumn{3}{c|}{$^{17}_\Lambda$O}& \multicolumn{3}{c}{$^{41}_\Lambda$Ca }\\
   \cline{2-10}
                                &  $1s_{1/2}$  &  $1p_{3/2}$ &  $1p_{1/2}$ &  $1s_{1/2}$ &  $1p_{3/2}$ &  $1p_{1/2}$&  $1s_{1/2}$ &  $1p_{3/2}$ &  $1p_{1/2}$     \\
  \hline

     Schmidt                       &  -0.613& -0.613  & 0.204  & -0.613  & -0.613 &  0.204   & -0.613           & -0.613            &   0.204 \\
     Relativistic                  &  -0.651& -0.644  & 0.184  & -0.660  & -0.662 &  0.170   & -0.682          &  -0.718           &   0.153\\
     Tensor*                       &  -0.611& -0.636  & 0.195  & -0.611  & -0.646 &  0.186   & -0.611           &  -0.670           &   0.168    \\
     Tensor                        &  -0.611& -0.632  & 0.194  & -0.611  & -0.643 &  0.186   & -0.611           &  -0.667           &   0.170         \\
  \hline
  \hline
 \end{tabular}
 \label{tab1}
 \end{table}
  \begin{table}[h!]
   \centering
   \tabcolsep=8pt
   \caption{Polarized Dirac magnetic moments of $\Lambda$-hypernuclei
   $^{13}_\Lambda$C, $^{17}_\Lambda$O, $^{41}_\Lambda$Ca in units of
   nuclear magnetons (n.m.). The notations are the same as those in Table~\ref{tab1}. }
   \begin{tabular}{ c |c c c|ccc|ccc}
   \hline
   \hline
   \multirow{2}*{$\mu_D$ ($\times10^{-4}$ n.m.)}   & \multicolumn{3}{c|}{$^{13}_\Lambda$C} &   \multicolumn{3}{c|}{$^{17}_\Lambda$O}& \multicolumn{3}{c}{$^{41}_\Lambda$Ca }\\
   \cline{2-10}
                                &  $1s_{1/2}$  &  $1p_{3/2}$ &  $1p_{1/2}$ &  $1s_{1/2}$ &  $1p_{3/2}$ &  $1p_{1/2}$&  $1s_{1/2}$ &  $1p_{3/2}$ &  $1p_{1/2}$     \\
  \hline

     Relativistic      & -395  & -322  & -215       & -483   &  -503 &  -369  & -701          &   -1065           &   -546    \\
     Tensor*           &  1.08 & -241  & -113       & 1.25   &  -346 &  -207  & 1.40          &   -580            &   -390     \\
     Tensor            &  1.12 & -198  & -124       & 1.29   &  -311 &  -203  & 1.51          &   -555            &   -372    \\
  \hline
   \hline
 \end{tabular}
 \label{tab2}
 \end{table}

\begin{figure}[ht]
  \centering
  \includegraphics[width=8cm]{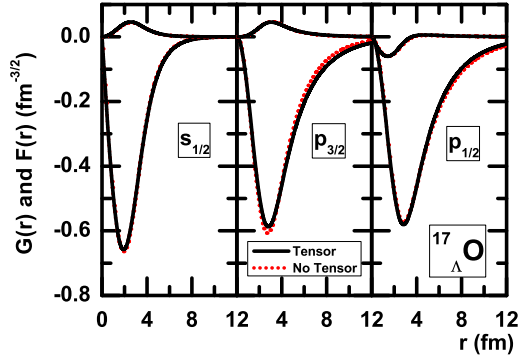}
   \caption{The $G(r)$ and $F(r)$ components in the Dirac spinors for the
   $1s_{1/2},1p_{3/2},1p_{1/2}$ $\Lambda$ hyperon states
   in $^{17}_\Lambda$O. The solid (dotted)
   lines are obtained with (without) the tensor potential.}
 \label{fig1}
 \end{figure}

 In Table~\ref{tab1}, it is seen that there is a large difference
 between the Relativistic and the Schmidt magnetic moment. Such a difference
 is mainly due to the polarized Dirac magnetic moments
 shown in Table~\ref{tab2}.  After taking into account the tensor
 effect on the current via Eq.~(\ref{PolarizedCT}), the difference
 almost disappears for the $\Lambda$ hyperon in the $1s$ state. Although
 this difference is greatly suppressed, there is still a $5\%-25\%$
 difference for a $\Lambda$ hyperon in the $1p$ state. Using the hyperon
 wave function obtained from the Dirac equation either without or with
 the $\Lambda$ tensor potential, the effect of $U^T_\Lambda$ on the
 magnetic moment has been studied as in Table~\ref{tab1}.
 Although the tensor potential is very important in reducing the
 spin-orbit splitting, its effect on the magnetic moment via the wave
 function is less than $0.1\%$. As shown in Fig.~\ref{fig1},
 for $1s_{1/2},1p_{3/2},1p_{1/2}$ states of a $\Lambda$ hyperon in
 $^{17}_\Lambda$O, the wave functions are almost the same irrespective
 of whether or not the $\Lambda$ tensor potential is included in the Dirac equation.
 Therefore the effect of the tensor potential on magnetic moments is
 small according to Eq.~(\ref{Tensor:DiracMM}).

\begin{figure}[ht]
  \centering
  \includegraphics[width=8cm]{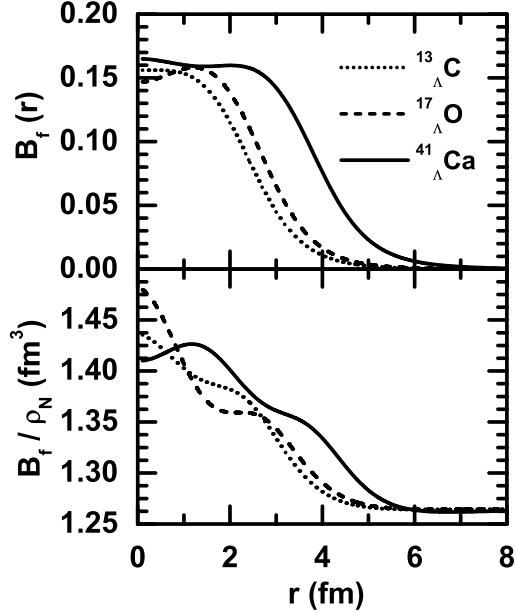}
   \caption{The reduction factor $B_f(r)$, defined by Eq. (\ref{Bfunction}) and
   the ratio $B_f/\rho_N$ for $^{13}_\Lambda$C, $^{17}_\Lambda$O, $^{41}_\Lambda$Ca,
   where $\rho_N(r)$ refers to the density distribution of the nuclear core.}
 \label{fig2}
 \end{figure}

 As shown in Table~\ref{tab1}, the differences with the Schmidt values
 for both $s_\Lambda$ and $p_\Lambda$ states increase with the mass
 number, which is due to the reduction factor $B_f(r)$ in Eq.~(\ref{Bfunction}).
 As $M^{\ast 2}_N \gg k_F^2$, one has
 $B_f(r)\propto[1+\dfra{M^\ast_N}{\lambda_{Nv}\rho_N(r)}]^{-1}$,
 i.e., large density distribution will lead to large $B_f(r)$.
 In Fig.~\ref{fig2}, the reduction factor $B_f(r)$ and the ratio $B_f/\rho_N$
 for $^{13}_\Lambda$C, $^{17}_\Lambda$O, $^{41}_\Lambda$Ca are given.
 The ratio $B_f/\rho_N$ decreases monotonically around 1.45~fm$^3$ to a constant
 value $1.26$~fm$^3$ for $\rho_N(r)\rightarrow 0$.
 Compared with $^{13}_\Lambda$C and $^{17}_\Lambda$O, the reduction factor $B_f(r)$
 for $^{41}_\Lambda$Ca is much larger, which results in a large difference from
 the Schmidt magnetic moment shown in Table~\ref{tab1}.

 In summary, core polarization and tensor coupling effects on the
 magnetic moments have been studied within a Dirac equation with
 scalar and vector potentials for $^{13}_\Lambda$C, $^{17}_\Lambda$O,
 $^{41}_\Lambda$Ca $\Lambda$-hypernuclei. The effect of core polarization
 on the magnetic moments is suppressed by $\Lambda$ tensor coupling.
 Although, the $\Lambda$ tensor potential reduces the
 spin-orbit splitting of $p_\Lambda$ states
 considerably, it has a negligible effect on the magnetic moments.
 The deviations of magnetic moments for $p_\Lambda$ states from the
 Schmidt values are found to increase with nuclear mass number.
 There is urgent need for experimental data to test these effects.

 It has to be pointed out that core polarization and tensor coupling effects on the
 magnetic moments of hypernuclei have been considered based only on the
 perturbation theory for the symmetric nuclear matter. A more self-consistent
 calculation in finite hypernuclei is required, in which all the nucleons and hyperon
 are treated on the same footing. The latter will be presented in a future paper.

\vspace{2em} This work is partly supported by Major State Basic
Research Developing Program 2007CB815000, the National Natural
Science Foundation of China under Grant Nos. 10435010, 10775004, and
10221003, the National Research Foundation of South Africa under
Grant No. 2054166, and the Scientific Research Foundation of Chinese
Agriculture University under Grant No. 2007005.


\end{document}